\documentclass[aps,prb,reprint,superscriptaddress]{revtex4-2}
\usepackage{amssymb}
\usepackage{amsmath}
\usepackage{booktabs}
\usepackage{siunitx}
\usepackage{mhchem}
\usepackage{tabularx}
\usepackage{wrapfig} 
\usepackage{siunitx}
\usepackage{mathtools}
\usepackage[margin=1in]{geometry}
\usepackage{float}
\begin{document}

\title{Optimized growth of large-size, high quality $\text{ZrTe}_5$ single crystals enabling clear quantum oscillations in electrical transport}

\author{Hong Du}
\thanks{These authors contributed equally to this work.}
\affiliation{Tsung-Dao Lee Institute, Shanghai Jiao Tong University, Shanghai 201210, China}

\author{Yu Cao}
\thanks{These authors contributed equally to this work.}
\affiliation{Tsung-Dao Lee Institute, Shanghai Jiao Tong University, Shanghai 201210, China}
\affiliation{School of Physics and Astronomy, Shanghai Jiao Tong University, Shanghai 200240, China}

\author{Jiahao Chen}
\affiliation{State Key Laboratory of Low Dimensional Quantum Physics, Department of Physics, Tsinghua University, 100084, Beijing, China}

\author{Tian Liang}
\affiliation{State Key Laboratory of Low Dimensional Quantum Physics, Department of Physics, Tsinghua University, 100084, Beijing, China}
\affiliation{Frontier Science Center for Quantum Information, 100084, Beijing, China}

\author{Liang Liu}
\affiliation{Tsung-Dao Lee Institute, Shanghai Jiao Tong University, Shanghai 201210, China}
\affiliation{School of Physics and Astronomy, Shanghai Jiao Tong University, Shanghai 200240, China}

\author{Ruidan Zhong}
\email{rzhong@sjtu.edu.cn}
\affiliation{Tsung-Dao Lee Institute, Shanghai Jiao Tong University, Shanghai 201210, China}
\affiliation{School of Physics and Astronomy, Shanghai Jiao Tong University, Shanghai 200240, China}

\date{\today}

\begin{abstract}
Quantum oscillation with nontrivial Berry phase is one of the characteristics of topological materials. As a Dirac semimetal candidate, zirconium pentatelluride ($\text{ZrTe}_5$) stands out as an intriguing material for investigating topological phase transitions and Dirac fermion physics; however, the extreme sensitivity of its electronic properties to stoichiometric variations and crystalline defects has hindered consistent experimental observation. Here, we report an optimized Te-flux synthesis method designed to produce centimeter-scale, high-quality single crystals meanwhile minimizing extrinsic carrier contamination. Comprehensive morphology, structural and chemical characterizations, including scanning electron microscopy, Laue backscattering and Rietveld refinement, confirm a high-purity $Cmcm$ phase with excellent crystallinity. Furthermore, magnetotransport measurements reveal a remarkably low Shubnikov-de Haas oscillation onset field ($B_{int} \approx 0.38$ T) with an ultra-high mobility of $5.58\times10^5$cm$^2$V$^{-1}$s$^{-1}$ and access to the the quantum limit at $B \approx 1.3$ T, attesting to the superior crystalline quality and the efficacy of this growth optimization. These results demonstrate that growth control is crucial for stabilizing intrinsic electronic behavior in $\text{ZrTe}_5$, establishing a robust platform for exploring topological phase transitions and exotic quantum phenomena in topological semimetals.
\end{abstract}

\maketitle

\section{Introduction}

Quantum oscillations in metallic systems originate from the quantization of the energy states into the Landau levels (LLs), when electrons are subjected to an external magnetic field $B$ and consequently forced into quantized cyclotron motion. As the magnetic field varies, the density of states at the Fermi level is modulated with a periodicity of $1/B$, leading to periodic oscillations in physical quantities. However, for two-dimensional (2D) or three-dimensional (3D) Dirac semimetals (DSMs) featuring topologically robust Dirac nodes with linear dispersion, the Landau quantization is fundamentally different due to the presence of relativistic fermions and nontrivial topology. This leads to exotic phenomena such as large linear magnetoresistance\cite{liang2015,shekhar2015}, chiral anomaly\cite{parameswaran2014,gorbar2014}, and anomalous quantum Hall effect\cite{nishihaya2025}, all of which are promising for future electronic and spintronic applications. Most strikingly, when all the electrons are condensed into the lowest LL—a regime known as the quantum limit—materials may exhibit new quantum states\cite{song2017,armitage2018,pan2019,sun2018,sun2025}. Reaching this limit requires the magnetic field to be comparable to the quantum oscillation frequency, necessitating either extremely high magnetic field\cite{zhao2015} or high-quality samples with low carrier density\cite{tangThreedimensionalQuantumHall2019,xingRashbasplittinginducedTopologicalFlat2024}.

$\text{ZrTe}_5$ was initially proposed as a candidate for a quantum spin Hall insulator in the 2D limit, with its bulk phase predicted to manifest as either a strong or weak topological insulator (TI) depending on the interlayer coupling\cite{weng2014a}. Despite these predictions, experiments investigations into bulk $\text{ZrTe}_5$ have yielded contradictory conclusions. Some studies reported $\text{ZrTe}_5$ as a weak TI with a global band gap by scanning tunneling microscopy and angle-resolved photoemission spectroscopy\cite{wu2016a,fan2017a,xiong2017a,zhang2017a}, while others suggested it is a 3D DSM through Shubnikov–de Hass (SdH) quantum oscillation\cite{zheng2016,yuQuantumOscillationsInteger2016,wang2018c,wang2018b}. This long-standing discrepancy has been largely reconciled by the consensus that $\text{ZrTe}_5$ sits in close proximity to topological phase boundary between a weak TI and a strong TI regime. At this critical transition point, the formation of a bulk Dirac node allows the system to effectively approximate a DSM state\cite{huTransportTopologicalSemimetals2019,tajkov2022,chenye2025}. As a result, the topology property of $\text{ZrTe}_5$ is highly sensitive to sample quality, which can be easily influenced by the growth methods, growth conditions and the measurement environment\cite{shahiBipolarConductionPossible2018,chen2018,mutchEvidenceStraintunedTopological2019,zhang2021a,galeskiOriginQuasiquantizedHall2021,gaikwad2022,song2022,salawu2023}. Systematic studies comparing samples grown via Te-flux and chemical vapor transport indicate that the flux method generally produces crystals closer to ideal stoichiometry with lower carrier density\cite{tangThreedimensionalQuantumHall2019,shahiBipolarConductionPossible2018}. Nevertheless, a closer inspection of the resistivity anomaly peak temperature ($T_p$)—which reflects the Fermi level position relative to the Dirac point—reveals that the Fermi surface varies significantly even among flux-grown samples. Specifically, flux-grown $\text{ZrTe}_5$ samples in this study\cite{tangThreedimensionalQuantumHall2019} exhibit resistivity peaks in the $60\text{--}100$ K range, while others display a persistent semiconducting behavior extending down to 2 K\cite{shahiBipolarConductionPossible2018}, suggesting a profound sensitivity of the electronic structure to subtle growth variations. Therefore, optimizing the growth of $\text{ZrTe}_5$ to produce reproducible, large-size, high-quality crystals is crucial not only for the study of quantum oscillation and reaching quantum limit, but also for comprehensively map the delicate topological properties of $\text{ZrTe}_5$.

In this paper, we report an optimized Te-flux growth method to synthesize large-size, high-quality $\text{ZrTe}_5$ single crystals with ultra-low carrier density and ultra-high mobility. By implementing a strategic thermal oscillation protocol and a crucible-reduced setup, we successfully suppressed the spontaneous nucleation and enlarged the resulting crystal dimensions to centimeter scale. Meanwhile, by utilizing purified Te and an alumina filter, we significantly eliminated extrinsic carrier contamination in $\text{ZrTe}_5$. The superior quality of these crystals is corroborated by comprehensive characterizations via SEM, Laue backscattering, and Rietveld refinement. Most significantly, magnetotransport measurements reveal the onset of SdH oscillations at a remarkably low magnetic field ($B_{int} \approx 0.38$ T) with a high mobility of $5.58\times10^5$cm$^2$V$^{-1}$s$^{-1}$, and the system entering the quantum limit by $B \approx 1.3$ T. These results not only validate the exceptional sample quality achieved through the optimized growth method, but also provide a robust platform to unravel the topological properties of $\text{ZrTe}_5$.

\section{Experimental details}

\subsection{Optimized sample growth method}

Large-scale single crystals of $\text{ZrTe}_5$ were synthesized using an optimized Te-flux method designed to maximize crystal dimensions while minimizing extrinsic carrier density. Given that Te is highly susceptible to oxidation in ambient air, a pre-growth purification was performed. As displayed in Figure 1(a), commercially available raw tellurium (6N purity) was vacuum-sealed with active carbon and annealed at $700^\circ\text{C}$ for 8 hours in a single-zone furnace. During annealing, the carbon reacts with and reduces tellurium oxides, while simultaneously absorbing volatile impurities. We found that temperatures above $700^\circ\text{C}$ are also effective for this purification process. Although direct characterization of the purified Te is not provided, single crystals grown from the purified Te exhibit significantly improved quality—characterized by lower carrier concentration and higher mobility—compared to those synthesized from as-purchased raw Te. This treatment effectively suppresses the formation of oxide inclusions and scattering centers. High-purity Zr (99.9999\%) and purified Te were utilized in a 1:300 molar ratio. To scale up production and increase crystal size, the starting material mass was increased from 30g to 60g by omitting an internal crucible, allowing the quartz tube itself to act as the primary growth vessel. Moreover, to prevent contamination from traditional quartz wool, which often introduces silicon-based impurities or creates unwanted nucleation sites, an alumina filter was integrated into the assembly by a pre-necking in the middle of the quartz tube, as illustrated in Figure 1(b)), providing a more stable and localized environment for single-crystal growth. The temperature profile of the growth process is presented in Figure 1(c). The sealed assembly was heated to $900^\circ\text{C}$ and held for 48 hours to ensure a fully homogenized melt. Following an initial rapid cool-down to $600^\circ\text{C}$ (150 min), the cooling profile was specifically engineered with strategic thermal oscillations to control nucleation and facilitate the growth of large-scale, high-perfection crystals. The melt was slowly ramped down to $469^\circ\text{C}$ at a rate of $1^\circ\text{C/h}$ to promote crystallization. This was followed by a swift reheating to $509^\circ\text{C}$ within 10 minutes and then slowly down to $469^\circ\text{C}$ at $1^\circ\text{C/h}$. This rapid remelt process is critical for dissolving smaller and defective nuclei into the flux allowing the solute to recrystallize onto larger, existing single crystals. After several repeated cycles, the furnace was heated to $490^\circ\text{C}$ in 5 minutes. The ampoule was then removed and centrifuged, successfully decanting the molten Te through the alumina filter into the catch crucible and isolating the large-size $\text{ZrTe}_5$ single crystals at the bottom of the tube in Figure 1(d).

\begin{figure*}
\includegraphics[width=1\textwidth]{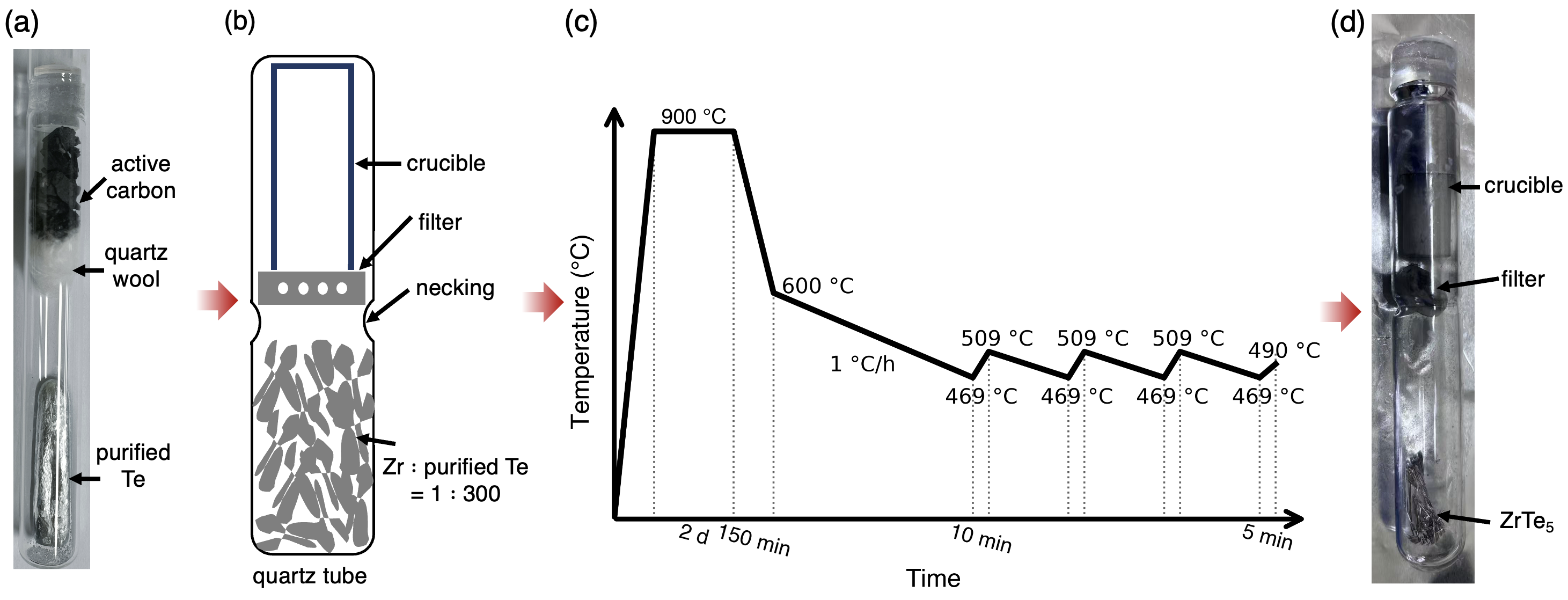}
\caption{(a) Purified Te by annealing at $700^\circ\text{C}$ with active carbon. (b) Schematic illustration of the growth ampoule assembly, featuring a quartz tube with one catch crucible and a necking design to support a specialized alumina filter. (c) Temperature profile of the optimized Te-flux growth process. (d) Photograph of the quartz tube after centrifugation at $490^\circ\text{C}$, showing the separation of the residual Te-flux from the harvested $\text{ZrTe}_5$ crystals.}
\label{fig1}
\end{figure*}

\subsection{Measurement methods}

The surface morphology and elemental analysis of the prepared single crystal samples were inspected by scanning electron microscopy (SEM) coupled with Energy Dispersive Spectroscopy (EDS) using ZEISS Gemini 300. The crystallinity of the grown crystals was examined by the Laue backscattering technique with a polychromatic source of the x-ray beam. The structural characterization was performed by Powder X-ray diffraction (PXRD), using a Bruker D8 Advance Eco powder diffractometer equipped with (Cu-$K_\alpha$ radiation) ($\lambda$ = 1.5418 \AA). Rietveld refinement were carried out on the powder diffraction data using the TOPAS v.6.0 software. Electrical connections are made using gold wires and silver paste. Both linear and non-linear electrical transport measurements are conducted using Quantum Design 14T Dynacool. For six-terminal linear transport measurements, longitudinal and transverse resistances are measured using a K6221 AC current source and SR830 lock-in amplifiers. For non-linear transport measurements, signals are monitored by the lock-in technique with harmonic number = 2, relative to the excitation current ($f$ = 13.777 Hz).

\section{Results and Discussion}

\subsection{Sample Morphology}

As shown in Figure 2(a\textendash b), by using the optimized Te-flux method with a thermal oscillation protocol, we successfully scaled the crystal dimensions by nearly an order of magnitude. Initially, we reproduced typical single crystals measured approximately $2 \times 0.2 \times 0.05$ mm$^3$ following growth protocols described in previous reports\cite{shahiBipolarConductionPossible2018}. However, by implementing our optimized Te-flux method we consistently attained large-scale crystals reaching up to $10 \times 2 \times 1$ mm$^3$. This substantial enlargement in all three dimensions confirms the effectiveness of the suppressed nucleation approach. The photograph of the as-grown $\text{ZrTe}_5$ crystal exhibit ribbon-like morphology with shiny metallic luster, indicating their high crystallinity and characteristic quasi-two-dimensional structure. Scanning electron microscopy (SEM) characterization shows the resulting crystals have well-defined crystal layers, which corresponds to the $ac$ plane of crystal structure, while characteristic quasi-one-dimensional (quasi-1D) stripes align with the $a$-direction (Figure 2(c)). As shown in the high-magnification micrograph of the crystal corner (Figure 2(d)), the individual layers are clearly stack along the $b$-axis, which is a direct manifestation of the unique crystal structure of $\text{ZrTe}_5$. These morphology features, combined with the absence of significant grain boundaries or defects, suggest the high sample quality required to observe clear quantum oscillations in transport measurements.

\begin{figure*}
\includegraphics[width=1\textwidth]{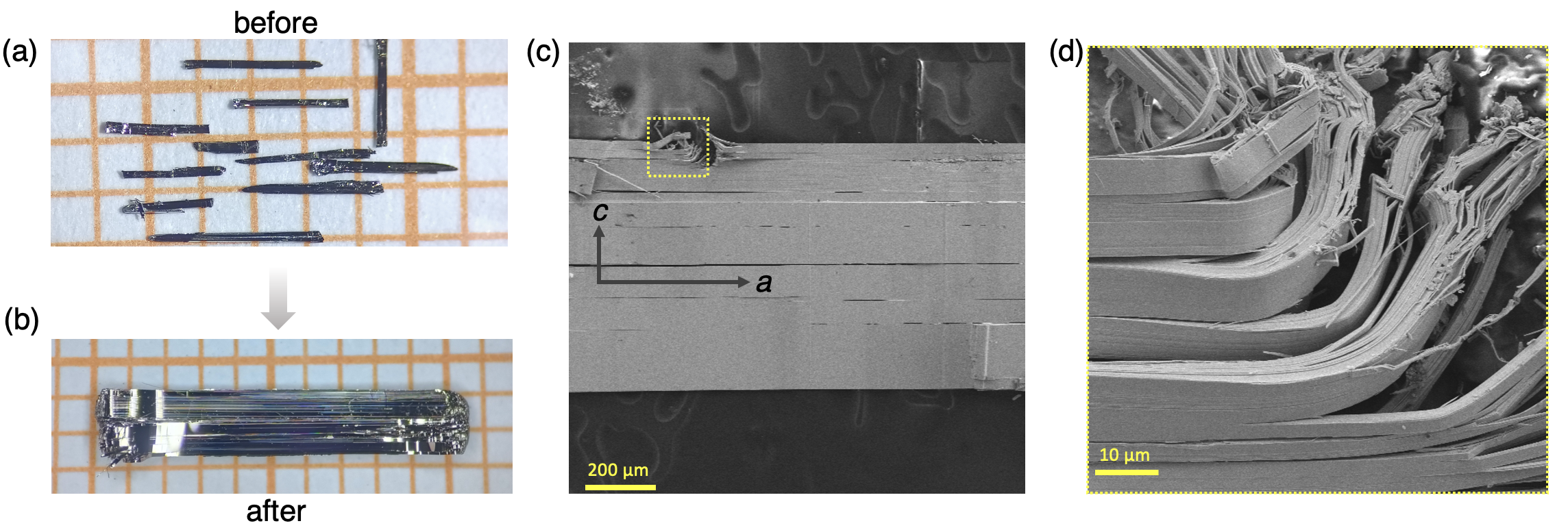}
\caption{(a\textendash b) Optical images of typical $\text{ZrTe}_5$ single crystals grown by previously reported method\cite{shahiBipolarConductionPossible2018} and the optimized Te-flux method, with (b) showing large-size, ribbon-like morphology with length exceeding 1 cm. (c) Scanning electron microscopy (SEM) image of a representative crystal, displaying a flat, high-quality surface. (d) High-magnification SEM image of the area highlighted in (c), revealing the distinct layered structure characteristic of $\text{ZrTe}_5$.}
\label{fig2}
\end{figure*}

\subsection{Structural and chemical characterization}

$\text{ZrTe}_5$ crystallizes in the orthorhombic layered crystal structure with space group $Cmcm$, as presented in Figure 3(a\textendash b). In the unit cell, one zirconium atom and three tellurium atoms are arranged into a $\text{ZrTe}_3$ tetrahedron. The $\text{ZrTe}_3$ tetrahedrons run parallel to the $a$-axis to form infinite quasi-1D chains, which are further linked along the $c$-axis via zigzag chains of Te atoms to form 2D layers in the $ac$ plane. These two-dimensional $\text{ZrTe}_5$ layers stack along the b-axis via weakly van der Waals forces with a half-period shift between two neighboring layers, forming the 3D bulk crystal. Hence, the single crystals are easy to mechanically exfoliate to thin flakes of $\text{ZrTe}_5$.

The single crystallinity of $\text{ZrTe}_5$ crystal was examined using the X-ray Laue backscattering measurement. The resulting Laue diffraction pattern, as shown in Figure 3(c), exhibits sharp and clear diffraction spots without any evidence of broadening or splitting, ascertaining the excellent crystal quality. The observed symmetry of the Laue diffraction pattern is in excellent agreement with theoretical calculations for the $Cmcm$ space group along (010) direction, verifying that the large-area facets of the crystal platelets are oriented along the $ac$ plane.

The obtained $\text{ZrTe}_5$ crystals were fully ground to identify the crystal phase by the powder X-ray diffraction (XRD) measurements and Rietveld refinement. As shown in Figure 3(d), all the diffraction peaks can be well fitted and no other phases are detected, confirming the phase purity of the compound. The calculated lattice constants are $a$ = 3.99 \AA, $b$ = 14.53 \AA, $c$ = 13.73 \AA and the cell volume is 795.1 \AA$^3$, which are in close agreement with the previously reported values\cite{shahiBipolarConductionPossible2018}. The refinement converges well with the reliability factors $R_{wp}$ = 8.79, reflecting a reliable fitting result. The detailed information about the structure refinement is given in Table I. Figure 3(e) shows the XRD results measured on the exposed surface of $\text{ZrTe}_5$ crystals. All the peaks can be indexed to $(0 k 0)$ planes, indicating that the exposed surface belongs to to $ac$ plane. The quality of the crystal was further examined using the X-ray rocking curve, as shown in the inset of Figure 3(e). The X-ray rocking curve of the (020) peak displays a full-width at half-maximum (FWHM) of only $0.06^\circ$, confirming the exceptional crystalline mosaicity of the synthesized samples. 

Elemental stoichiometry and sample homogeneity were verified by energy-dispersive X-ray spectroscopy (EDS) measurement. The representative spectrum of on numbers of freshly cleaved crystals is shown in Figure 3(f). The average $\ce{Zr}:\ce{Te}$ molar ratios are estimated to be $1:4.95$, which is near to the ideal stoichiometric ratio. The element mapping of Zr and Te, presented in Figure 3(g), shows the uniform distribution of Zr and Te, strongly suggesting that the growth is homogeneous in the synthesized samples. The above comprehensive structural characterizations, together with the elemental analysis rigorously demonstrate the high single-crystalline quality and phase purity achieved through the optimized Te-flux growth method.

\begin{figure*}
\centering
\includegraphics[width=1\textwidth]{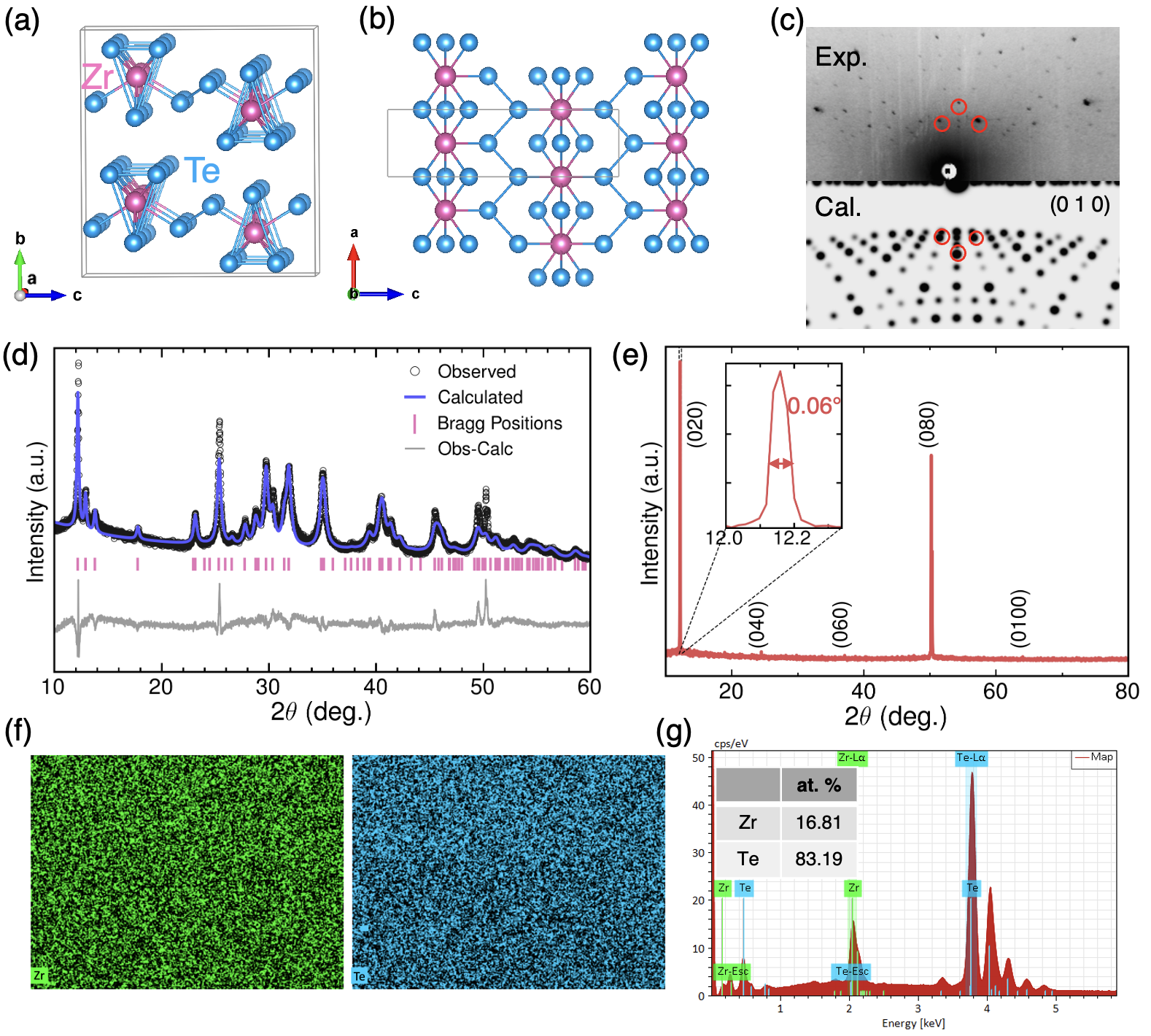}
\caption{(a\textendash b) Schematic illustration of the orthorhombic crystal structure of $\text{ZrTe}_5$ viewed along the $a$- and $b$-axes, highlighting the quasi-1D $\text{ZrTe}_3$ chains and the layered 2D structure. (c) Laue backscattering pattern of a flux-grown single crystal, showing sharp, well-defined spots that match the calculated pattern (bottom). (d) Powder X-ray diffraction (XRD) pattern and Rietveld refinement of crushed $\text{ZrTe}_5$ crystals. (e) XRD pattern of a $\text{ZrTe}_5$ single crystal showing $(0k0)$ reflections. Insets show a rocking curve of the $(020)$ peak with a FWHM of $0.06^\circ$. (f) EDS elemental mapping for Zr (green) and Te (blue), showing a uniform distribution of elements across the crystal. (g) EDS spectrum confirming the chemical composition, with an atomic ratio of approximately 1:4.95, indicating near-ideal stoichiometry.}
\label{fig3}
\end{figure*}

\begin{table*}
\caption{\label{tab:zrte5_full}Crystallographic data and structure refinement of $\text{ZrTe}_5$ single crystals at room temperature.}
\begin{ruledtabular}
\begin{tabular}{ll c llllll}
Refined Formula & \ce{ZrTe5} & & Atom & Wyckoff & $x$ & $y$ & $z$ & Occ.\\
\cline{1-2} \cline{4-9}
Crystal system & Orthorhombic & & Te(1) & $8f$ & 0.0000 & 0.07010 & 0.14940 & 1 \\
Space group & $Cmcm$ (No. 63) & & Te(2) & $8f$ & 0.0000 & 0.20990 & 0.56470 & 1 \\
$a$ (\si{\angstrom}) & 3.9866(4) & & Te(3) & $4e$ & 0.0000 & 0.33650 & 0.25000 & 1 \\
$b$ (\si{\angstrom}) & 14.5276(1) & & Zr & $4e$ & 0.0000 & 0.68570 & 0.25000 & 1 \\
\cline{4-9}
$c$ (\si{\angstrom}) & 13.7286(9) & & & & & & & \\
$V$ (\si{\angstrom^3}) & 795.11(2) & & & & & & & \\
Density (\si{g/cm^3}) & 6.092 & & Reliability & $R_{\text{exp}}$ (\%) & $R_{\text{wp}}$ (\%) & $R_{p}$ (\%) & \multicolumn{2}{l}{GOF} \\

Formula weight & 729.22 & & factor & 3.83 & 8.79 & 6.74 & \multicolumn{2}{l}{2.30} \\
\end{tabular}
\end{ruledtabular}
\end{table*}

\subsection{Transport measurement}

To verify the sample quality through electrical transport, the longitudinal resistivity ($\rho_{xx}$) and Hall resistivity ($\rho_{xy}$) were measured at 1.9 K. As illustrated in Figure 4(a), the zero-field resistivity exhibits a characteristic increase upon cooling, peaking at a temperature $T_p = 90$ K. This peak is attributed to a temperature-induced Lifshitz transition, where the Fermi level shifts across the Dirac point, leading to a crossover in the dominant carrier species\cite{liChiralmagneticeffect2016}. The high quality of the optimized crystals is most prominently evidenced by the observation of pronounced Shubnikov-de Haas (SdH) quantum oscillations. Figure 4(b) shows the longitudinal resistivity ($\rho_{xx}$) as the functions of the perpendicular magnetic field ($B$, applied along the $b$-axis). More than three sets of SdH oscillations in $\rho_{xx}$ can be well observed below 2 T with a remarkably low onset field of $B_{int} \approx$ 0.38 T, which indicates the extremely high mobility in our flux-grown crystals. Hall mobility and carrier concentrations are obtained by low-magnetic-field Hall measurements (Fig. 4(c)), from which we extract the density $n=9.20\times10^{16}$ cm$^{-3}$ and mobility $\mu=5.58\times10^5$ cm$^2$V$^{-1}$s$^{-1}$. Notably, a kink is observed in the magnetoresistance near $B\approx$ 1.3 T, signifying the system's entry into the quantum limit. This feature persists when the magnetic field is tilted from the out-of-plane direction ($b$-axis) toward the in-plane direction ($c$-axis), suggesting the robustness of the quantum limit in this 3D Dirac semimetal.

\begin{figure*}
\centering
\includegraphics[width=1\textwidth]{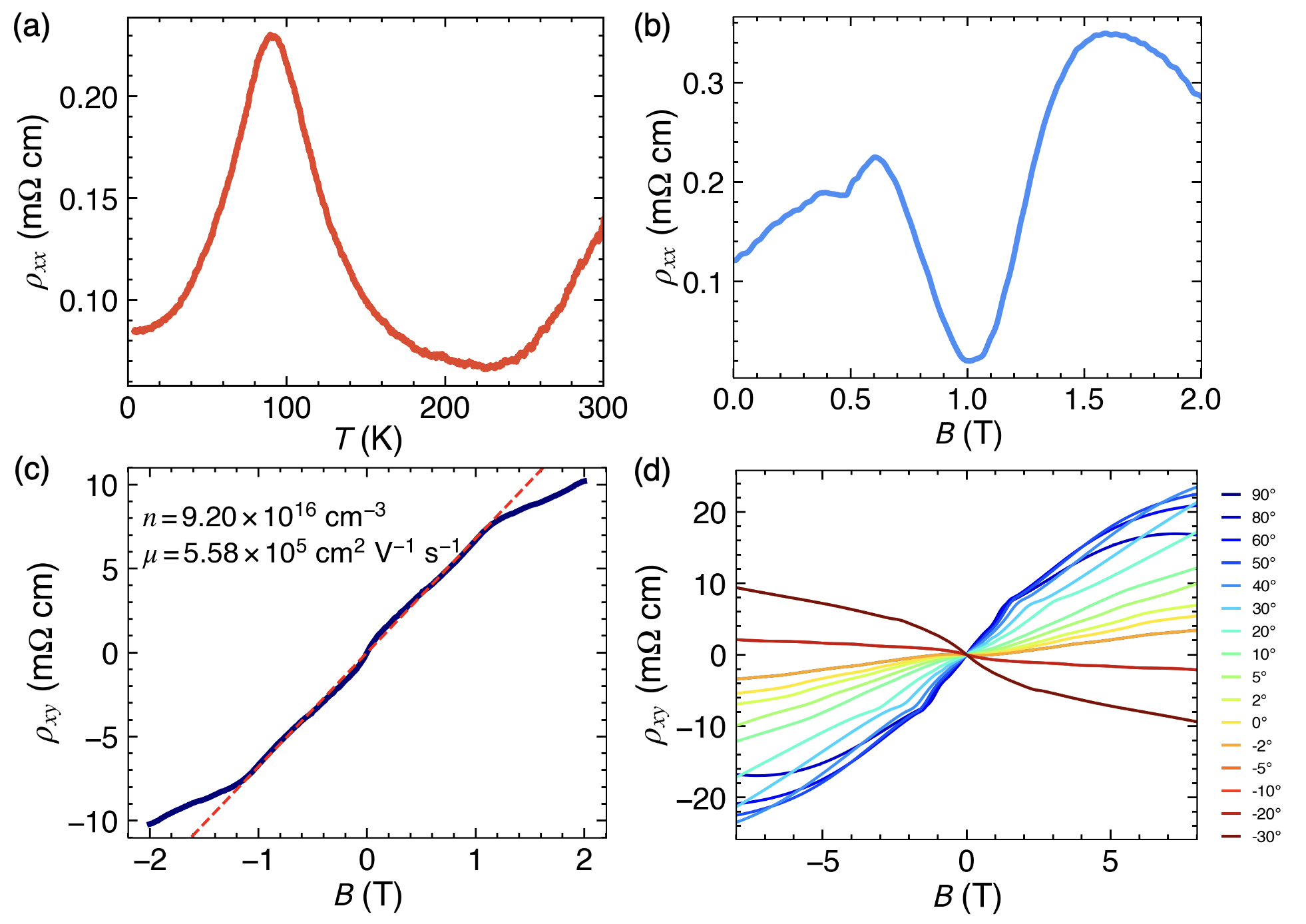}
\caption{(a) Temperature dependence of the longitudinal resistivity $\rho_{xx}$, showing the characteristic resistivity anomaly peak at $T_p \approx 90$ K. (b) Low-temperature $\rho_{xx}$ versus magnetic field $\mu_0 H$, exhibiting more than three clear Shubnikov-de Haas (SdH) oscillations with remarkably low onset fields ($B_{int} \approx 0.38$ T), confirming high mobility. (c) Hall resistivity $\rho_{xy}$ at $T = 1.9$ K. The dashed line represents a linear fit for $|B| < 0.6$ T, from which an ultra-low carrier density ($n = 9.20 \times 10^{16}$ cm$^{-3}$) and ultra-high mobility ($\mu = 5.58 \times 10^5$ cm$^2$ V$^{-1}$ s$^{-1}$) are extracted. (d) Angle-dependent $\rho_{xy}$ curves as the magnetic field is rotated from the $b$-axis to the $ac$ plane.}
\label{fig4}
\end{figure*}

\section{Conclusion}
\label{con}
In conclusion, we have demonstrated an optimized Te-flux synthesis technique that effectively suppresses the spontaneous nucleation and eliminates extrinsic carrier contamination in $\text{ZrTe}_5$. By the strategic use of purified precursor, specialized growth assembly and a thermal oscillation profile, this method enables the growth of centimeter-scale single crystals with exceptional structural perfection. Comprehensive characterizations via SEM, Laue backscattering, and Rietveld refinement confirm the synthesis of a high-purity $Cmcm$ phase. The high crystalline quality is further validated by the emergence of Shubnikov-de Haas oscillations at a remarkably low onset field of 0.38 T with an ultra-high mobility of $5.58\times10^5$cm$^2$V$^{-1}$s$^{-1}$. Additionally, the observation of a distinct Hall resistivity signature at $B \approx 1.3$ T signifies the system’s entry into the quantum limit, substantiating the efficacy of this growth optimization. This work provides a controllable methodology for obtaining high-quality $\text{ZrTe}_5$, establishing a robust platform for further probing the sensitive topological phase transitions and exotic quantum phenomena inherent to this Dirac semimetal candidate.

\begin{acknowledgments}
The work was supported by the National Key R\&D Program of China under 2021YFA1401600 and 2022YFA1402702, National Natural Science Foundation of China with Grants Nos. 12334008, and 12374148. The experimental support from the Instrumental Analysis Center at Shanghai Jiao Tong University, especially in performing the EDS analyses.
\end{acknowledgments}

\bibliography{ZT5}

\end{document}